# How Visa-Free Policies Fuel International Research Collaboration: Evidence from China

Songlin Cai[1], Xuan Liu[2], Xianwen Wang[1*]

1. School of Public Administration and Policy, Dalian University of Technology, Dalian 116024, China
2. National Academy of Innovation Strategy, China Association for Science and Technology, Beijing 100038, China

* Corresponding author. Email address: xianwenwang{[at]}dlut.edu.cn

**Abstract**

Visa regimes constitute significant institutional barriers to the cross-border mobility of researchers. Utilizing China's phased implementation of a unilateral visa-free policy since 2023 as a quasi-natural experiment, this study employs a staggered difference-in-differences design to assess the policy's effect on international scientific collaboration. Results indicate that the policy significantly increased the volume of Sino-foreign co-authored publications. The mechanism analysis indicates that this effect is primarily achieved by enhancing transportation accessibility and human mobility, which in turn facilitates cross-border research collaboration among scholars. Further evidence suggests that academic conferences partially attenuated the policy's impact, indicating a substitutive relationship across collaboration channels. Moreover, the effect was more pronounced for countries with greater geographical distance or lower research capacity. This study elucidates the mechanisms through which visa facilitation promotes international scientific collaboration and offers new insights into how institutional barriers shape research cooperation and knowledge production.

## 1. Introduction

International research collaboration has become a major driver of knowledge production and technological advancement (Adams, 2013). Endogenous growth theory emphasizes that the diffusion and sharing of knowledge continuously stimulate innovation, thereby generating sustained economic growth (Aghion and Howitt, 1992). In recent years, cross-border academic collaboration has expanded rapidly (Gök and Karaulova, 2024; Thelwall et al., 2024), most visibly through the sharp increase in internationally coauthored publications (Isfandyari-Moghaddam et al., 2023). This trend is reflected not only in the growth of research output but also in the intensified exchanges across broader geographic and institutional boundaries (Xu and Zong, 2025). The contemporary landscape of innovation research is inherently collaborative, and scientific cooperation has emerged as a global phenomenon (Robinson et al., 2023). Driven jointly by the Belt and Road Initiative and the strategy of research internationalization, China's exchanges and cooperation with countries around the world have continued to deepen, and its international research collaboration networks have grown increasingly close. In 2023, internationally coauthored papers accounted for 20.4% of all publications from China, involving partners from 177 countries and regions. Such collaboration often relies on frequent site visits and face-to-face interactions, which raise higher demands on time, funding, transportation, and institutional environments. However, non-academic factors such as visa regimes, geopolitical tensions, and public health crises create structural barriers to international research collaboration by restricting researcher mobility and impeding knowledge exchange, thereby constraining the scope and depth of global collaboration. Against this backdrop, China's recent phased implementation of a unilateral visa-free policy provides a new institutional opportunity to reduce barriers to cross-border exchanges and improve conditions for researcher mobility, while offering a unique setting to evaluate its impact on international research collaboration.

Research collaboration has become a core mode of knowledge production and innovation (Adams, 2013, Zhao et al., 2022). At the international level, coauthorship networks have continued to expand, with cross-border publications accounting for a growing share of research output (Gazni et al., 2012). Existing studies attribute this trend to several factors, including advances in communication technologies (Agrawal and Goldfarb, 2008; Wernsdorf et al., 2022), increased cross-border mobility of researchers (Franzoni et al., 2014; Scellato et al., 2015; Gu et al., 2024), improvements in research funding and policy environments (Crescenzi and Rodríguez-Pose, 2017), general population policy settings (Woolley et al., 2015), and the structural effects of academic networks (Hennemann et al., 2012). In addition, geographic proximity and transportation infrastructure have been shown to play a critical role in shaping the patterns of research collaboration (Hoekman et al., 2010; Catalini, 2018; Kang et al., 2023).

However, in the context of rapidly expanding research systems in developing countries—particularly China—the impact of visa regimes on international research collaboration has yet to be systematically examined. Existing studies have primarily focused on exogenous factors such as transportation, communication, and research funding, while discussions of visa policies remain limited and largely centered on bilateral visa-waiver agreements (Czaika and Parsons, 2017). Research on unilateral visa-free policies is still scarce. Unlike reciprocal visa waivers or short-term transit visa exemptions, unilateral visa-free arrangements reflect a host country's proactive openness. Such policies may reduce collaboration barriers by facilitating the cross-border mobility and face-to-face interactions of researchers, thereby promoting knowledge exchange and international research collaboration. Against this backdrop, this study investigates whether and how unilateral visa-free policies promote collaboration between Chinese and foreign scholars by easing researcher mobility and communication. Specifically, we address two core questions: first, do unilateral visa-free policies significantly enhance the international mobility and communication of scientists? Second, does this institutional innovation further strengthen research collaboration between Chinese and foreign scholars?

This study evaluates the impact of the unilateral visa-free policy on research collaboration between Chinese and foreign scholars. To this end, we employ a staggered difference-in-differences design, exploiting institutional variation in the timing at which different countries were added to the visa-free list. We compare changes in the number of Sino-foreign coauthored publications before and after policy implementation in treated countries with those in unaffected countries. The empirical results show that the unilateral visa-free policy significantly increased the number of coauthored papers, confirming its positive role in fostering international research collaboration. Mechanism analysis further shows that the policy facilitates international research collaboration by improving transportation accessibility and promoting human mobility. Moreover, when academic exchanges had already occurred through conferences, the marginal effect of the visa-free policy diminished, suggesting a partial substitution between collaboration channels. Heterogeneity analyses reveal that the policy's facilitative effect on research collaboration was more pronounced for countries located at greater geographic distances and those with relatively weaker research capacity.

A central challenge in our DID design lies in addressing potential endogeneity. Specifically, the selection of countries into the visa-free list may not be random, as government decisions could be influenced by unobserved factors related to research collaboration, such as bilateral relations, trade linkages, or scientific capacity (Czaika and Parsons, 2017). Without proper controls, these factors could bias the estimates. To mitigate this concern, we first examine the correlation between the visa-free policy and countries' levels of technological innovation, finding no evidence that innovation capacity determined inclusion in the list. Second, we include countries that had been listed for the visa-free policy but had not yet implemented it during the study period in the control group, thereby improving both the comparability and external validity of the control group (Goodman-Bacon, 2021; Callaway and Sant'Anna, 2021). In addition, we exclude Belt and Road countries and those already enjoying reciprocal visa waivers to account for potential confounding effects of geopolitics and other institutional arrangements. Across these tests, our main findings remain robust.

In addition, we conducted a series of robustness checks to strengthen the credibility of our conclusions. First, we applied the sensitivity analysis framework of Rambachan and Roth (2023) to address the limitations of event-study designs in testing parallel trends. Second, we employed the stacked DID method proposed by Cengiz et al. (2019) to mitigate potential identification biases of two-way fixed-effects models under staggered adoption (De Chaisemartin and d'Haultfoeuille, 2020). Third, we performed a placebo test, which showed that the actual estimates significantly deviated from the placebo distribution, thereby reinforcing the causal interpretation. Fourth, we combined propensity score matching (Heckman et al., 1997) with DID to enhance the comparability between treatment and control groups. Fifth, we adopted the two-stage DID model of Gardner (2022) to correct for potential bias arising from heterogeneous treatment effects. Sixth, we implemented the synthetic DID (SDID) approach of Arkhangelsky et al. (2021), which relies on a data-driven synthetic control group to produce robust estimates. Finally, we conducted an alternative specification by replacing the dependent variable with Sino-foreign coauthored papers on bioRxiv, and the results remained consistent. Overall, these robustness checks confirm that the positive impact of the unilateral visa-free policy on Sino-foreign research collaboration is highly robust.

This paper is closely related to recent research on scientific collaboration and the facilitation of academic exchanges. For instance, Wernsdorf et al. (2022) show that the early academic network BITNET significantly increased the volume of collaborations but had limited effects on research quality, underscoring the constraints of online communication. In terms of offline interactions, Chen et al. (2025) find that the establishment of direct flights between cities substantially promoted intercity collaborative innovation, as reduced face-to-face communication costs enhanced knowledge flows. Similarly, Zhang et al. (2025), using subway construction as a quasi-natural experiment, demonstrate that urban rail development fostered research collaboration and, compared with online communication, contributed more strongly to research quality. While these studies focus on improvements in digital communication or physical transportation, they overlook the potential role of institutional barriers—such as visa policies—in shaping cross-border academic collaboration. On the other hand, Czaika and Parsons (2017) and Bahar et al. (2020) establish the link between high-skilled migration policies and international talent flows, but they do not directly examine the impact of visa facilitation on research output. Compared with this literature, our contribution lies in identifying the unilateral visa-free policy as an institutional shock. Such policies reflect a proactive stance of the host country and possess externalities that differentiate them from conventional exogenous factors like transportation infrastructure or communication technologies. By focusing on institutional facilitation, this study uncovers the mechanisms through which cross-border exchanges shape research collaboration, providing new empirical evidence on the institutional drivers of collaboration.

This study makes several marginal contributions across academic fields.

First, it links institutional visa facilitation to research collaboration, offering a new perspective on how cross-border exchange costs shape scientific cooperation. While prior studies have extensively examined the role of information and communication technologies or transportation infrastructure in lowering collaboration costs (Agrawal and Goldfarb, 2008; Wernsdorf et al., 2022; Dong et al., 2020; Koh et al., 2025), the influence of visa regimes has received little direct attention. In particular, existing work on high-skilled mobility has largely focused on bilateral waivers or broader immigration policies (Czaika and Parsons, 2017; Bahar et al., 2020), whereas this study identifies the causal effects of a unilateral visa-free policy. Such a policy is more proactive and empirically identifiable, reducing entry barriers while creating a broader institutional environment for international academic collaboration.

Second, this study examines the short-term effects of the unilateral visa-free policy using a more fine-grained temporal scale. Most existing research relies on annual data, which makes it difficult to identify the immediate impact of the policy on the dynamics of research collaboration. By using monthly data, this paper not only improves the timeliness and sensitivity of effect identification but also provides a new perspective for understanding the dynamic adjustment of international research collaboration under exogenous institutional shocks.

Finally, this paper innovatively employs preprint data to measure collaboration. Prior studies have predominantly relied on published papers, a practice with notable limitations: first, the formal publication date typically lags behind the actual initiation of collaboration, making it difficult to capture the true timing of research partnerships; second, published papers represent only a subset of scientific output, overlooking research that never reaches formal publication channels. Figure. A1 illustrates the difference between preprints and traditional publication pathways. In contrast, preprints record submission dates, allowing for a more immediate reflection of collaboration onset, while also covering both published and unpublished work. This offers a more comprehensive view of international research collaboration. Given that preprints have become important channels of scholarly exchange in many disciplines (e.g., arXiv in mathematics and computer science, bioRxiv in biology), our use of preprint data not only alleviates the problems of publication lag and incomplete coverage but also enriches the toolkit for measuring international research collaboration.

The remainder of this paper is organized as follows. Section 2 introduces the theoretical framework underlying the empirical analysis. Section 3 outlines the background of China's unilateral visa-free policy. Section 4 describes the data collection procedures, data sources, and empirical strategy in detail. Section 5 presents the estimation results and provides an in-depth discussion of the core research questions. Section 6 concludes.

## 2. Theoretical framework

Offline interactions and face-to-face exchanges are critical for fostering scientific collaboration and knowledge production (Boschma, 2005; Ulnicane, 2015; Catalini, 2018). However, visa restrictions, as institutional barriers, may hinder cross-border mobility (Czaika and Neumayer, 2017) and thereby limit researchers' opportunities to attend academic conferences, undertake short-term visits, or pursue joint projects. Against this backdrop, unilateral visa-free policies mitigate visa-related barriers and substantially reduce the institutional costs of international mobility, greatly facilitating short-term visits to China. Existing research further highlights that face-to-face interaction is not only a channel for information exchange but also a central mechanism for building trust, transmitting tacit knowledge, and stimulating innovative ideas (Uzzi et al., 2013; Wernsdorf et al., 2022). International research collaboration often depends on direct personal contacts established during conferences, visiting appointments, or fieldwork, which have been shown to significantly increase the likelihood of cross-border coauthorship and to broaden and deepen international collaboration networks (Franzoni et al., 2014; Freeman and Huang, 2015). By enabling such interactions, visa-free policies serve as a vital bridge for exchanges between Chinese and foreign scholars, accelerating the transfer and sharing of knowledge and effectively promoting international research collaboration. Accordingly, we propose:

**Hypothesis 1.** Unilateral visa-free policies significantly promote international research collaboration between China and visa-exempt countries.

*2.1. Transportation accessibility and international research collaboration*

Beyond mobility barriers, the formation and maintenance of research collaboration are also shaped by the level of transportation accessibility. The extent of high-speed rail coverage and the degree of air connectivity directly affect the financial and time burden researchers must bear to engage in collaboration (Catalini et al., 2020). Moreover, the effective organization of activities such as academic conferences, fieldwork, and joint experiments relies heavily on efficient and convenient transportation. Prior studies show that improvements in travel speed and accessibility significantly enhance cross-regional and international collaboration among high-skilled workers, as reduced travel frictions expand the geographic reach of researchers and broaden the pool of potential collaborators (Catalini et al., 2020). From the perspective of innovation network formation, research collaboration often requires frequent interaction and sustained communication, and greater accessibility facilitates more frequent contacts, thereby improving the efficiency of knowledge spillovers and innovation diffusion. By stimulating demand and supply in bilateral air transport, unilateral visa-free policies encourage airlines to increase flight frequency, optimize transfers, and even launch direct routes, thereby improving transportation accessibility for foreign scholars traveling to China for collaboration. Enhanced accessibility not only strengthens researchers' willingness to cooperate but also improves the efficiency and outcomes of joint projects. Based on this reasoning, we propose:

**Hypothesis 2.** Unilateral visa-free policies promote international research collaboration by improving transportation accessibility.

*2.2. Human mobility and international research collaboration*

International research collaboration is strongly constrained by limitations in human mobility, with visa requirements serving as explicit barriers that significantly impede cross-border academic exchange. Prior studies demonstrate that visa restrictions substantially reduce international mobility, thereby weakening the potential for cross-country collaboration (Czaika and Neumayer, 2017). For researchers, visa applications not only impose direct economic and time costs but also entail considerable administrative complexity and uncertainty (Gülel, 2025). Such barriers dampen the willingness and feasibility of attending international conferences, engaging in short-term visits, or pursuing joint research projects, ultimately hindering the formation and deepening of high-quality collaborations (Appelt et al., 2015). From the perspective of international knowledge flows, mobility constraints distort the efficient cross-border allocation of resources, restricting knowledge spillovers and the organic growth of collaboration networks (Aman, 2022). Removing these barriers can enhance researcher flows, increase opportunities for face-to-face interactions, and facilitate more effective collaboration matching (Liu and Hu, 2022; Han et al., 2024). Accordingly, unilateral visa-free policies expand the feasible frontier of international collaboration by improving human mobility, thereby fostering international research collaboration. Based on this reasoning, we propose:

**Hypothesis 3.** Unilateral visa-free policies promote international research collaboration by enhancing human mobility.

*2.3. Substitution effects across collaboration channels*

Research collaboration typically relies on multiple channels, including formal institutional facilitation measures (e.g., visa policies, research funding) as well as informal academic exchange platforms (e.g., international conferences, scholarly communities). Conferences and workshops not only serve as key venues for presenting research but also provide crucial opportunities for building trust and expanding collaboration networks (Freeman and Huang, 2015). Face-to-face interaction plays an irreplaceable role in promoting information sharing and knowledge spillovers, helping to reduce search costs and cognitive frictions between potential collaborators (Boschma, 2005; Uzzi et al., 2013). However, when different collaboration channels perform similar functions in lowering cross-border exchange costs and fostering opportunities, substitution effects may emerge—that is, the strengthening of one channel can diminish the marginal contribution of another. For example, geographic proximity and institutional facilitation partially overlap in promoting collaboration (Boschma, 2005; Catalini, 2018), while the rise of online communication tools has reduced, to some extent, the need for face-to-face interaction (Brucks and Levav, 2022). Consequently, when one channel (e.g., academic conferences) already provides opportunities for collaboration, the marginal effect of another channel (e.g., visa-free policies) may weaken (Freeman and Huang, 2015; Boudreau et al., 2017). Based on this reasoning, we propose:

**Hypothesis 4.** The substitution effects across collaboration channels weaken the positive impact of unilateral visa-free policies on international research collaboration.

**3. Background**

Many countries and international organizations have adopted visa waivers and facilitation measures to promote cross-border mobility, stabilize trade, investment, and tourism, and consolidate their competitive advantages in the global economy. For example, since the Schengen Agreement took effect in 1995, the European Union has eliminated border visa checks among member states, facilitating the free movement of people and significantly advancing regional economic integration. In a similar vein, the United Arab Emirates signed a reciprocal short-term visa waiver agreement with the EU in 2015, proactively easing entry restrictions for EU citizens to strengthen its role as a hub for international exchange and commerce.

Amid the progress of the Belt and Road Initiative and the growing demand for cross-border mobility, China's traditional visa regime has increasingly revealed constraints in terms of time, financial costs, and uncertainty, limiting the further optimization of its openness strategy. To address these challenges, China began phasing in a unilateral visa-free policy in 2023, aiming to enhance the efficiency of cross-border mobility, expand openness, and reinforce its hub position in global economic networks.

China first piloted visa-free entry for 59 countries in Hainan Province in 2018, providing valuable experience for subsequent nationwide expansion. Starting in 2023, the policy was broadly implemented, initially covering six countries—France, Germany, Italy, the Netherlands, Spain, and Malaysia—allowing ordinary passport holders to enter China visa-free for up to 15 days for business, tourism, family visits, or transit. In 2024, the list was further extended to include Switzerland, Ireland, Hungary, Austria, Belgium, and Luxembourg, with the maximum length of stay also increased. As shown in Table 1, by June 9, 2025, the unilateral visa-free policy had expanded to 47 countries across Asia, Europe, and South America.

The policy has already yielded substantial benefits. In 2024, China recorded 20.115 million visa-free entries through national ports of entry, a year-on-year increase of 112.3%. Of these, 3.391 million foreign nationals entered under the unilateral visa-free policy, marking a 1200.6% increase and accounting for 13.8% of all inbound foreign entries (excluding border residents). These figures highlight not only the policy's direct impact in expanding inbound flows but also its role in reducing institutional costs, thereby enhancing the convenience and predictability of cross-border mobility and laying an institutional foundation for deeper international cooperation.

The inclusion of countries on the visa-free list is primarily determined by macro-strategic considerations such as diplomatic relations, consular agreements, and immigration risk, rather than by economic or research performance. As a result, the risk of reverse causality is relatively low when evaluating the policy's impact on research collaboration, providing favorable identification conditions. Nevertheless, although the policy itself is not designed to directly promote research collaboration, it remains necessary to consider the potential endogeneity arising from a country's technological capacity influencing its likelihood of being added to the list. To address this concern, we use publication counts and patent applications as proxy variables for scientific research and technological innovation capacity. The empirical results (Table A.1) show no significant effects, indicating that the formulation of the visa-free policy is not systematically driven by technological performance. This finding alleviates, at least in part, concerns about endogeneity bias stemming from country-level external factors.

**Table 1**

Implementation process of China's unilateral visa-free policy.

| Batch | Starting date | Newly added visa-free countries | Total |
|---|---|---|---|
| - | 2023.7.26 | Brunei, Singapore | 1 |
| 1 | 2023.12.1 | France, Germany, Italy, Spain, the Netherlands, Malaysia | 7 |
| 2 | 2024.3.14 | Switzerland, Ireland, Hungary, Austria, Belgium, Luxembourg | 13 |
| 3 | 2024.7.1 | New Zealand, Australia, Poland | 16 |
| 4 | 2024.10.15 | Portugal, Greece, Cyprus, Slovenia | 20 |
| 5 | 2024.11.8 | Slovakia, Norway, Finland, Denmark, Iceland, Andorra, Monaco, Liechtenstein, Republic of Korea | 29 |
| 6 | 2024.11.30 | Bulgaria, Romania, Croatia, Montenegro, North Macedonia, Malta, Estonia, Latvia, Japan | 38 |
| 7 | 2025.6.1 | Brazil, Argentina, Chile, Peru, Uruguay | 43 |
| 8 | 2025.6.9 | Saudi Arabia, Oman, Kuwait, Bahrain | 47 |

Note: Since 2003, China had granted a 15-day unilateral visa-free entry to Brunei, Singapore, and Japan, but this policy was suspended on March 28, 2020. On July 26, 2023, China resumed the unilateral visa-free arrangement for Brunei and Singapore. On February 9, 2024, Singapore introduced a reciprocal visa-free policy for Chinese citizens, and thus it is not included in the definition of unilateral visa-free countries in this paper. On November 30, 2024,

China reinstated unilateral visa-free entry for Japan and, from that date, extended the maximum stay from 15 to 30 days for all 38 eligible countries. Source: Compiled by the authors.

## 4. Data and empirical strategy

*4.1. Data*

**International Research Collaboration.** To measure the level of international research collaboration, this study relies on preprint data, which are particularly suitable for capturing collaboration dynamics in a timely and comprehensive manner. Compared with formally published papers, preprints record submission dates and are released without lengthy peer-review and production lags, thereby providing a more accurate measure of the actual timing of collaborative activities (Larivière et al., 2014). Moreover, preprints include both eventually published and unpublished outputs, thus offering broader coverage of scientific collaboration than publication-based datasets. Within this framework, we use the number of coauthored papers between Chinese and foreign scholars published on the arXiv platform as the core indicator. ArXiv is the world's largest preprint server, currently hosting more than two million scholarly articles across eight major fields, including physics, mathematics, and computer science, and has been widely recognized for its representativeness in the study of scientific collaboration. To ensure robustness, we also validate our results using publication data from the bioRxiv platform, which has become an increasingly important repository for life sciences. The dataset spans July 2023 to November 2024, fully covering the policy implementation period of interest, and thus aligns with the quasi-experimental design of this study. Since the number of collaborations is zero for certain countries in some months, we apply the inverse hyperbolic sine (IHS) transformation to the coauthorship counts, which allows us to mitigate zero-value bias while preserving sample continuity.

**Unilateral Visa-Free Policy.** Given the time lag in publication data, we define the treatment group as the 15 countries included in the first three batches of China's unilateral visa-free policy between December 2023 and July 2024. These countries are France, Germany, Italy, Spain, the Netherlands, Malaysia, Switzerland, Ireland, Hungary, Austria, Belgium, Luxembourg, New Zealand, Australia, and Poland. They span multiple continents and vary considerably in academic capacity, making them well-suited for evaluating policy effects. For the control group, we draw on the Web of Science database to identify all countries with which China had co-authored papers on the arXiv platform during 2021–2022. We then exclude countries already covered by China's unilateral visa-free policy, yielding a final control group of 66 countries.

**Control Variables.** At the country level, we include four control variables to better account for potential confounding factors. First, we use the monthly number of invention patent applications (*Patents*) as a proxy for national innovation capacity, with data obtained from the IncoPat database. IncoPat is one of the most comprehensive international patent information platforms, covering over 190 million patent records from 170+ countries, regions, and organizations worldwide. Second, to capture overall research performance, we measure each country's publication output on the arXiv preprint platform (*Preprints*) and in the Web of Science Core Collection (*Journals*). Given publication lags, both variables are included with a one-period lag. Finally, we construct a geographic accessibility index (*Accessibility*) to reflect the ease of travel to China, defined as 1 / (geographic distance to China × aviation fuel price index), where higher values indicate greater accessibility. Descriptive statistics for selected variables are shown in Table 2.

**Table 2**

Descriptive statistics.

| Variable | # of Obs. | Mean | SD | Min | Median | Max |
|---|---|---|---|---|---|---|
| Papers | 1377 | 25.393 | 89.258 | 0 | 1 | 1048 |
| Visa-free | 1377 | 0.102 | 0.303 | 0 | 0 | 1 |
| Patents | 1377 | 1012.687 | 2820.084 | 0 | 42 | 29839 |
| Preprints | 1377 | 251.047 | 771.833 | 0 | 17 | 8761 |
| Journals | 1377 | 2429.633 | 6153.662 | 8 | 512 | 59760 |
| Accessibility | 1377 | 0.006 | 0.004 | 0.002 | 0.004 | 0.043 |

*4.2. Empirical strategy*

To examine the impact of the visa-free policy on international research collaboration, we construct the following difference-in-differences (DID) model, as shown in Equation (1):

$$Papers_{it} = \alpha + \beta \times Visa\_free_{it} + \gamma X_{it} + \mu_i + \upsilon_t + \varepsilon_{it} \qquad (1)$$

where $Papers_{it}$ is the dependent variable, measured as the number of arXiv coauthored papers between country $i$ and China in month $t$, transformed using the inverse hyperbolic sine function. $Visa\_free_{it}$ is the core explanatory variable, defined as the interaction of $treat_i$ and $post_t$. Specifically, $treat_i = 1$ if country$_i$ is included in China's unilateral visa-free list and 0 otherwise, while $post_t = 1$ for months after the policy took effect for that country and 0 otherwise. $X_{it}$ is a vector of control variables capturing other determinants of international research collaboration, including innovation capacity, research capacity, and travel accessibility. $\mu_i$ and $\upsilon_t$ denote country fixed effects and month fixed effects, which account for time-invariant unobserved heterogeneity at the country level (e.g., geographic proximity) and common temporal shocks, respectively. $\varepsilon_{it}$ is the error term, clustered at the country level.

To further examine the dynamic effects of the policy and test the validity of the parallel-trends assumption, we employ an event-study specification as shown in Equation (2):

$$Papers_{it} = \alpha + \sum_{k=-5}^{4} \beta_k \times Visa\_free_{it}^k + \gamma X_{it} + \mu_i + \upsilon_t + \varepsilon_{it} \qquad (2)$$

where $Visa\_free_{it}^k$ is a binary indicator equal to 1 if country $i$ is in the visa-free list at month $t$, with $k$ denoting the relative month to the policy implementation ($k \in [-5, 4]$). Other variables are defined in the same way as in Equation (1).

## 5. Results

*5.1. Baseline results*

We begin by testing the effect of the visa-free policy based on Equation (1). Table 3 reports the estimation results. Column (1) shows that, compared with countries not covered by the policy, those included in China's unilateral visa-free list experienced a significant increase in the number of collaborative papers with China, indicating a positive effect of the policy on fostering international research collaboration. In Column (2), after including control variables, the estimated coefficient remains significant. To mitigate potential bias associated with count data, Column (3) uses the logarithm of paper counts, and the results remain significantly positive. In Column (4), we expand the control group to include countries that were listed for the policy but had not yet implemented it during the study period, and the estimated coefficient continues to be positive and statistically significant, suggesting that our core conclusion does not depend on a specific choice of control group. Taken together, the baseline regression results consistently support Hypothesis 1.

Next, we estimate the dynamic effects of the visa-free policy using Equation (2), with the year prior to each policy implementation batch serving as the reference period. The results are illustrated in Figure 1. Before policy implementation, the difference in paper counts between the treatment and control groups was small and statistically insignificant, consistent with the parallel-trends assumption. After implementation, the number of collaborative papers in the treatment group rose significantly, with positive and significant coefficients in both the first and fourth post-policy periods. Overall, the results reveal a sustained upward trajectory, indicating that the visa-free policy has a significant and lasting effect on promoting international research collaboration.

**Table 3**

The impact of the visa-free policy on international research collaboration.

|  | (1) Papers | (2) Papers | (3) Papers(log) | (4) Papers |
|---|---|---|---|---|
| Visa-free | 0.123** | 0.134*** | 0.128*** | 0.110** |
|  | (0.048) | (0.047) | (0.041) | (0.044) |
| Control variables | N | Y | Y | Y |
| Country FE | Y | Y | Y | Y |
| Month FE | Y | Y | Y | Y |
| Observations | 1377 | 1377 | 1377 | 1904 |
| R-squared | 0.951 | 0.951 | 0.959 | 0.946 |

Note: The table reports the effects of the visa-free policy on the number of international research collaboration papers. All specifications include country and month fixed effects. Column (2) extends the baseline model by adding control variables, Column (3) further uses the logarithmic form of the dependent variable, and Column (4) expands the control group to include countries that were listed for visa exemption during the study period but had not yet implemented the policy. ***, **, * indicate significance at the 1%, 5%, and 10% levels, respectively. Standard errors are reported in parentheses.

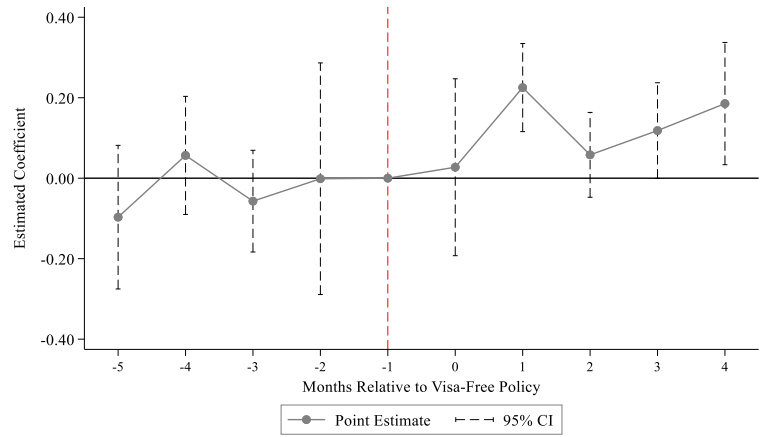

**Fig. 1.** Dynamic effect of the visa-free policy on collaborative publications
Note: This figure illustrates the dynamic impact of China's unilateral visa-free policy on the number of coauthored papers between Chinese and foreign scholars. The horizontal axis represents time relative to policy implementation (measured in months), while the vertical axis shows the estimated coefficients. Solid circles denote point estimates, and dashed lines represent the corresponding 95% confidence intervals. The coefficients remain close to zero prior to policy implementation, indicating that the parallel-trends assumption is largely satisfied. After implementation, the coefficients rise gradually and reach statistical significance in some periods, suggesting that the visa-free policy has, to some extent, promoted collaborative exchanges between Chinese and foreign scholars.

To further explore the heterogeneous effects of the unilateral visa-free policy, we conduct subgroup regressions based on geographic distance to China's capital and countries' research capacity. As reported in Table 4, the policy significantly increases the number of Sino-foreign coauthored papers in countries located farther from China, with a larger coefficient (0.176, $p < 0.01$), while the effect is insignificant for geographically closer countries. This suggests that the policy plays a particularly important role in overcoming spatial barriers to collaboration. In addition, using the number of core publications in the WOS database to measure research capacity, we find that the policy effect is significantly positive in countries with weaker research capacity (0.243, $p < 0.01$), but not significant in stronger research systems. This indicates that the visa-free policy generates greater marginal benefits by lowering barriers to collaboration and facilitating engagement for countries with more limited academic resources. Overall, the heterogeneity analysis demonstrates that the policy's positive impact is more pronounced for geographically distant and less research-intensive countries.

**Table 4**

Heterogeneity analysis.

|  | (1) Distant countries | (2) Nearby countries | (3) Strong research capacity | (4) Weak research capacity |
|---|---|---|---|---|
| Visa-free | 0.176*** | 0.064 | 0.085 | 0.243*** |
|  | (0.052) | (0.116) | (0.056) | (0.030) |
| Control variables | Y | Y | Y | Y |
| Country FE | Y | Y | Y | Y |
| Month FE | Y | Y | Y | Y |
| Observations | 714 | 663 | 731 | 646 |
| R-squared | 0.972 | 0.915 | 0.946 | 0.604 |

Note: This table tests the heterogeneous effects of the unilateral visa-free policy across different country characteristics. Geographic distance is measured by the straight-line distance to China's capital, and research capacity is proxied by the number of core publications in the WOS database. The results show that the policy effect is more pronounced in countries that are geographically distant from China and in those with weaker research capacity. ***, **, * indicate significance at the 1%, 5%, and 10% levels, respectively. Standard errors are reported in parentheses.

*5.2. Robustness check*

**Sensitivity Analysis.** Since pre-trend tests cannot serve as strict proof of the parallel-trends assumption, we further assess the robustness of the baseline results by applying the sensitivity analysis framework proposed by Rambachan and Roth (2023). The core idea of this method is to allow for bounded deviations from the strict parallel-trends assumption and evaluate the stability of the estimated effects under varying degrees of violations. Figure 2 presents the sensitivity analysis for the first and fourth post-policy periods. The horizontal axis represents the relative magnitude of allowed deviations from parallel trends, while the vertical axis shows the 95% confidence intervals of the estimated policy effects. The results indicate that in both the first and fourth periods, the confidence intervals remain entirely above zero, even when substantial deviations from parallel trends are permitted. This suggests that our baseline estimates are not driven by overly strong assumptions of strict parallel trends and remain robust under more flexible conditions. Therefore, the conclusion that the unilateral visa-free policy significantly promotes international research collaboration is both robust and credible.

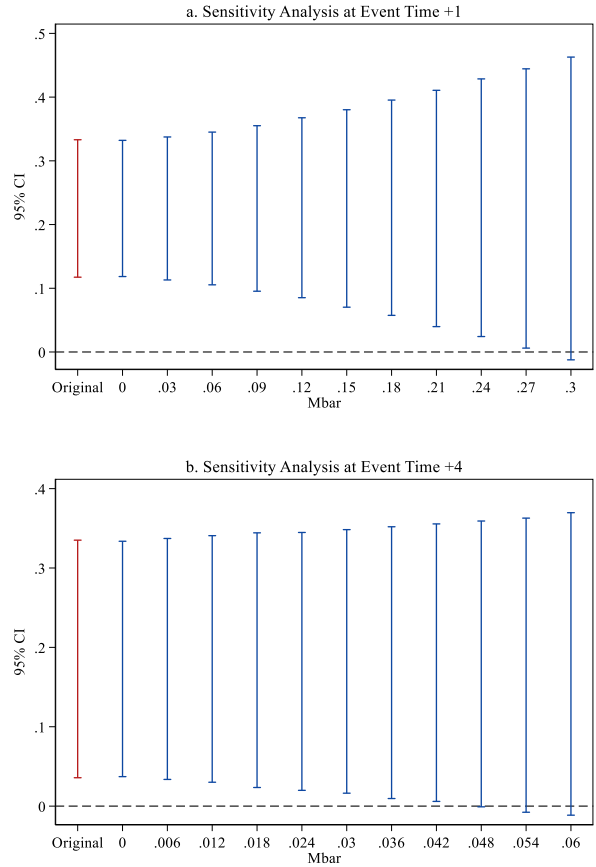

**Fig. 2.** Sensitivity test of the parallel-trends assumption
Note: This figure illustrates the 95% confidence intervals of the estimated treatment effects under varying degrees of deviation from the parallel-trends assumption. The horizontal axis denotes the maximum allowed proportionate deviation of post-treatment trends relative to pre-treatment trends, while the vertical axis shows the corresponding confidence interval range. If the confidence interval does not include zero for a given deviation level, the estimated policy effect remains valid under more relaxed assumptions, thereby enhancing the credibility and robustness of the causal inference.

**Stacked DID.** When estimating with a staggered DID design, variation in treatment timing may render the conventional two-way fixed-effects estimator invalid due to identification bias (Goodman-Bacon, 2021). Specifically, when treatment effects are heterogeneous, early-treated units may be inappropriately used as controls for later-treated units, leading to negative weighting issues (De Chaisemartin and d'Haultfoeuille, 2020). To mitigate this concern and further test robustness, we employ a stacked DID approach. This method re-aligns samples according to their respective treatment dates and estimates effects separately, thereby avoiding invalid comparisons under staggered adoption. As shown in Figure 3, the results from the stacked DID are consistent with the baseline estimates in both magnitude and direction, confirming that our core finding—that the visa-free policy promotes research collaboration—remains robust even in the presence of treatment-effect heterogeneity.

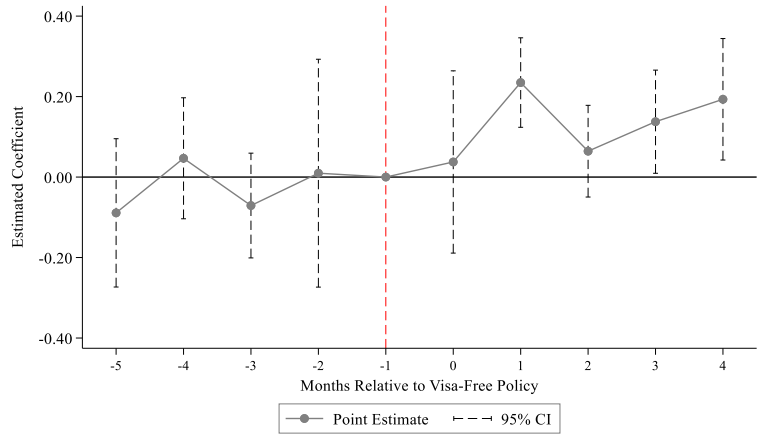

**Fig. 3.** Results of stacked DID estimation
Note: This figure presents the dynamic effect estimates based on the stacked DID method. The horizontal axis represents months relative to the implementation of the visa-free policy, and the vertical axis plots the estimated coefficients. Dots indicate point estimates, while dashed lines denote the 95% confidence intervals. The coefficients remain close to zero prior to policy implementation, supporting the parallel-trends assumption. After implementation, the coefficients rise gradually and become significant in some periods, indicating that the visa-free policy has a positive effect on promoting international research collaboration.

**Placebo Test.** At the country level, we conduct a stratified random sampling procedure to generate placebo treatment dates and re-estimate the treatment effects and their standard errors. This process is repeated 1,000 times. The results are presented in Figure 4, where the black solid line depicts the kernel density of the placebo estimates, the blue dots represent the simulated p-values, and the red dashed line indicates the baseline estimate. The majority of placebo estimates cluster around zero, whereas the actual estimate lies significantly in the right tail of the empirical distribution. This suggests that the policy effect identified in the baseline regression is unlikely to be driven by random chance.

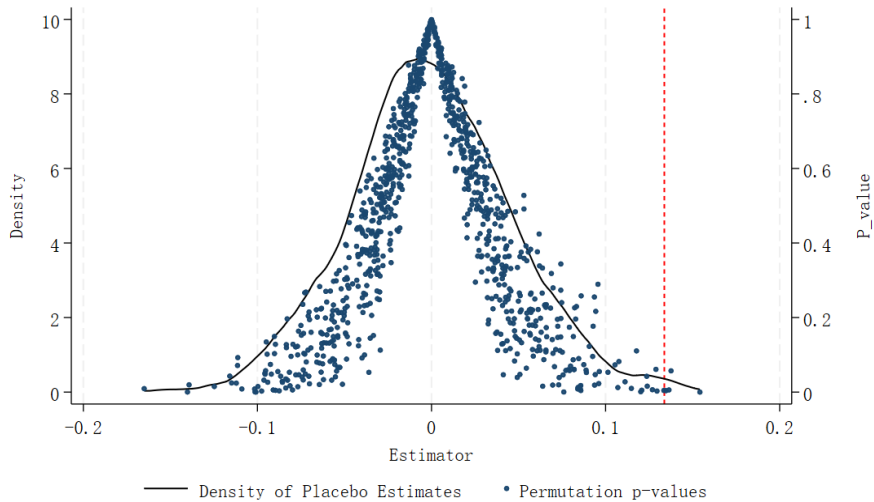

**Fig. 4.** Results of the placebo test
Note: This figure presents the results of the placebo test. The black solid line depicts the kernel density of the permutation estimates, the blue dots represent the simulated p-values, and the red dashed line indicates the actual estimated effect. The results show that the actual effect deviates significantly from the placebo distribution, confirming the robustness of the baseline regression results.

**PSM-DID.** To address potential selection bias arising from systematic differences in observable characteristics between the treatment and control groups, we conduct a robustness check by combining propensity score matching (PSM) with the DID approach. PSM is commonly used to control for confounders and to estimate the causal effect of treatments on the outcomes (Xu and Zong, 2024). Using both radius matching and kernel matching, we construct comparable control groups for each treated country and re-estimate the policy effect. The results, reported in Columns (1) and (2) of Table 5, show positive and statistically significant coefficients at the 5% level, indicating that the estimated treatment effect remains robust after accounting for sample characteristics.

**Two-Stage DID.** To account for potential issues of "invalid comparisons" and negative weights in staggered treatment settings, we implement the two-stage DID approach proposed by Gardner (2022) as a robustness check. In the first stage, we purge systematic variation in the outcome variable by controlling for country and month fixed effects along with a set of covariates. In the second stage, we regress the residuals to identify the policy effect. The results, reported in Column (3), show that the coefficient of the unilateral visa-free policy is 0.156 and statistically significant at the 1% level. This indicates that even after accounting for treatment-effect heterogeneity, the policy effect remains significantly positive, thereby reinforcing the robustness of our conclusions.

**Synthetic DID.** To further verify the robustness of the baseline results, we employ the synthetic difference-in-differences (SDID) method proposed by Arkhangelsky et al. (2021). This approach constructs an optimally weighted synthetic control group and combines it with time differencing to account for common shocks, thereby mitigating potential biases in the conventional DID framework. Specifically, we apply both the optimized and projected weighting schemes for covariates and use bootstrap procedures to compute standard errors. The results, reported in Columns (4) and (5), yield estimated coefficients of 0.159 and 0.160, both highly significant. These findings demonstrate that even under the stricter synthetic control framework, the positive effect of the unilateral visa-free policy remains robust.

**Alternative Dependent Variable.** As an additional robustness check, we replace the dependent variable from the number of arXiv publications to the number of bioRxiv publications, with data obtained from the OpenAlex database to address the limited coverage of this preprint platform in Web of Science. The results, reported in Column (6), show a policy effect of 0.134, statistically significant at the 1% level. This finding indicates that the positive impact of the unilateral visa-free policy on research collaboration is not confined to a specific preprint platform or disciplinary domain, thereby further reinforcing the robustness and generalizability of the baseline conclusion.

**Controlling for Other Factors.** To rule out potential confounding effects from other policy arrangements, we exclude from the sample both the Belt and Road Initiative (BRI) countries and those with reciprocal visa-waiver agreements with China. The results, reported in Columns (7) and (8), yield estimated coefficients of 0.163 and 0.110, both statistically significant at the 5% level. These findings indicate that even after accounting for broader cooperation mechanisms such as the BRI and reciprocal visa policies, the unilateral visa-free policy continues to exert a significant positive effect on Sino-foreign research collaboration.

**Table 5**

Robustness checks.

| | (1) Radius matching | (2) Kernel matching | (3) Two-stage DID | (4) SDID (optimized) | (5) SDID (projected) | (6) bioRxiv papers | (7) Excluding BRI countries | (8) Excluding visa-free countries |
|---|---|---|---|---|---|---|---|---|
| Visa-free | 0.118** | 0.129*** | 0.156*** | 0.159** | 0.160*** | 0.134*** | 0.163*** | 0.110** |
| | (0.053) | (0.048) | (0.047) | (0.065) | (0.062) | (0.047) | (0.036) | (0.051) |
| Control variables | Y | Y | Y | Y | Y | Y | Y | Y |
| Country FE | Y | Y | Y | Y | Y | Y | Y | Y |
| Month FE | Y | Y | Y | Y | Y | Y | Y | Y |
| Observations | 857 | 1366 | 1377 | 1377 | 1377 | 1377 | 748 | 1020 |

|  | | | | | | |
|---|---|---|---|---|---|---|
| R-squared | 0.948 | 0.950 | | 0.951 | 0.980 | 0.969 |

Note: This table reports a series of robustness checks for the baseline results. Columns (1) and (2) apply PSM-DID using radius matching and kernel matching, respectively, to mitigate sample selection bias. Column (3) adopts the two-stage DID method proposed by Gardner (2022) to address potential biases from treatment-effect heterogeneity in staggered adoption designs. Columns (4) and (5) employ the synthetic difference-in-differences (SDID) approach of Arkhangelsky et al. (2021). Column (6) replaces the dependent variable from the number of arXiv publications to the number of bioRxiv publications, testing the sensitivity of results to alternative measures of research output. Columns (7) and (8) exclude Belt and Road Initiative countries and countries with reciprocal visa-waiver agreements with China, respectively. Across all models, the estimated coefficients remain positive and statistically significant, further confirming the robustness of the core conclusion. ***, **, and * denote significance at the 1%, 5%, and 10% levels, respectively. Standard errors are reported in parentheses.

*5.3. Mechanism*

Face-to-face academic interactions are indispensable for research collaboration, as they not only facilitate efficient information exchange and trust-building but also substantially enhance the efficiency and output of joint research (Storper and Venables, 2004). By enhancing transportation accessibility and researcher mobility, the visa-free policy provides critical support for international research collaboration.

The empirical results are presented in Table 6. The results in Column (1) show that the implementation of the policy significantly increased the number of international flights to China, indicating that the policy intervention effectively stimulated the supply capacity of the air transport market and thereby further improved the connectivity and convenience of the international transportation network. The results in Column (2) indicate that the implementation of the policy significantly increased the number of foreign entrants to China. This finding suggests that by optimizing cross-border mobility conditions, the visa-free policy effectively expanded the scale of international human flows, thereby creating a broader human resource base and more opportunities for academic exchange to support international research collaboration. In sum, Hypotheses 2 and 3 are supported, indicating that the unilateral visa-free policy has promoted international research collaboration by enhancing transportation accessibility and expanding inbound human mobility.

The moderating effect reported in Column (3) indicates that collaboration opportunities realized through academic conferences negatively moderate the impact of the visa-free policy, Hypothesis 4 is supported. In other words, when part of international collaboration has already been achieved via conference interactions, the marginal effect of visa facilitation weakens. A plausible explanation is that, compared with journal publications, conference papers are bound by strict submission deadlines, which make collaborators more likely to rely on efficient online or hybrid meetings rather than time-consuming long-distance travel. As a result, academic conferences may substitute for some of the offline exchanges enabled by visa-free entry. This interpretation is consistent with the substitution-effect theory: once collaboration channels such as online meetings are actively used, the incremental benefits of additional facilitation policies diminish. For example, Yang et al. (2022), using evidence from Microsoft's remote collaboration data, demonstrate that once existing communication structures are strengthened, the marginal benefits of additional collaboration channels (e.g., online meetings) decline substantially.

Overall, the visa-free policy promotes international research collaboration by improving transportation accessibility and strengthening human mobility, thereby reducing barriers to cross-border exchanges. At the same time, substitution effects across collaboration channels suggest that the impact of the policy may be constrained by the presence of alternative forms of cooperation. This evidence sheds light on the complex relationship between facilitation policies and academic collaboration behaviors.

**Table 6**

Mechanism tests.

|  | (1) Transportation accessibility | (2) Human mobility | (3) Papers |
|---|---|---|---|
| Visa-free | 0.278** | 1700.322** | 0.119** |
|  | (0.131) | (792.458) | (0.049) |
| Conference papers |  |  | -0.000*** |
|  |  |  | (0.000) |
| Interaction term    |   |   | -0.002* |
|                     |   |   | (0.001) |
| Control variables   | Y | Y | Y       |
| Country FE          | Y | Y | Y       |
| Month FE            | Y | Y | Y       |
| Observations        | 1377 | 1377 | 1377 |
| R-squared           | 0.976 | 0.775 | 0.951 |

Note: Transportation accessibility and Human mobility denote the number of direct flights to China and the number of inbound foreign visitors per country per month, respectively. Conference papers refer to the number of coauthored conference papers between Chinese and foreign scholars by month. The Interaction term is constructed as the product of Visa-free and Conference papers. Columns (1) and (2) show that the policy greatly promotes international research collaboration by increasing flight frequency and enhancing human mobility across borders. Column (3) demonstrates that academic conferences exert a negative moderating effect on the policy impact, suggesting potential substitution effects across collaboration channels. ***, **, * indicate significance at the 1%, 5%, and 10% levels, respectively. Standard errors are reported in parentheses.

## 6. Conclusion and policy implication

### 6.1. Conclusion

International research collaboration has become a key driver of knowledge creation and technological progress. How to lower the costs of cross-border exchanges and expand opportunities for scholarly cooperation has long been a central question in the literature. Existing studies have largely focused on the role of transportation infrastructure and information and communication technologies in fostering collaboration. For example, the expansion of high-speed rail networks significantly reduces the spatial and temporal costs for researchers (Catalini et al., 2020), while the diffusion of the Internet enhances academic output by lowering search frictions and communication barriers (Forman and van Zeebroeck, 2012; Wernsdorf et al., 2022). Collectively, this line of research establishes a causal link between declining exchange costs and strengthened collaborative ties. However, the impact of institutional barriers on research collaboration has yet to be systematically explored. Drawing on China's phased implementation of a unilateral visa-free policy in recent years, this paper employs a staggered DID design to evaluate the effects of visa facilitation on international research collaboration. The findings are as follows:

First, we find that the unilateral visa-free policy significantly increased the number of collaborations between China and visa-exempt countries. This result is robust across multiple specifications, confirming the positive role of lowering institutional barriers to mobility in promoting scholarly cooperation. Moreover, the effect is more pronounced for countries located farther from China and those with relatively weaker research capacity. Second, the policy operates mainly through two channels: on the one hand, it improves flight connectivity between China and partner countries, thereby reducing transportation barriers to cross-border exchanges among researchers; on the other hand, it substantially promotes the inflow of foreign visitors to China, expanding the scale of academic exchanges and thus fostering international research collaboration. Finally, the presence of academic conferences as an alternative channel partly offsets the policy effect, suggesting a potential substitution across different forms of collaboration.

Although this paper provides a systematic assessment of the impact of the unilateral visa-free policy on international research collaboration, several limitations remain. First, the primary indicator used in this study is the number of co-authored papers. While this measure captures formal collaboration between scholars, it does not account for broader forms of research interaction, such as joint patent applications, collaborative research projects, or informal academic exchange networks. Second, this study does not address the policy's effect on the quality of research collaboration. Because the preprint data employed here mainly reflect publication counts and lack quality dimensions such as peer review outcomes or journal rankings, it is not possible to systematically evaluate the policy's impact on research quality. Future research should incorporate more granular data on researcher mobility and diversify the indicators of research output, thereby providing a more comprehensive understanding of the true effects of visa facilitation policies.

*6.2. Policy implication*

This study demonstrates that China's unilateral visa-free policy exerts a significant positive effect on fostering research collaboration between Chinese and foreign scholars. By lowering institutional barriers to cross-border mobility, the policy effectively increases the frequency of academic exchanges and collaboration opportunities. Accordingly, as China advances toward a higher level of openness, it would be advisable to expand the scope of visa-free countries while simultaneously streamlining entry procedures and improving border management to create a more convenient and predictable environment for academic exchange. Flight connectivity and increases in inbound researcher flows serve as key channels through which the policy operates, implying that visa exemptions alone are insufficient to fully unlock collaboration potential. Complementary measures—such as strengthening international flight supply and improving customs efficiency—are needed to systematically reduce the economic and time costs of cross-border mobility and thereby maximize the institutional benefits of the visa-free policy. In addition, the finding that academic conference collaboration partially substitutes for the policy effect suggests that different collaboration channels may compete at the margin. Thus, policies to promote international academic exchanges should emphasize complementarity and coordination between visa facilitation and other collaboration channels, avoiding duplicative allocation of resources. Finally, the stronger policy effects observed for countries with weaker research capacity or greater geographic distance highlight that the policy not only deepens collaboration with advanced economies but also creates new opportunities to expand scientific ties with developing countries. Looking ahead, visa facilitation should be combined with measures such as increased research funding, international talent programs, and joint cross-border grants, forming a comprehensive policy package to advance high-quality research collaboration and the globalization of knowledge creation.

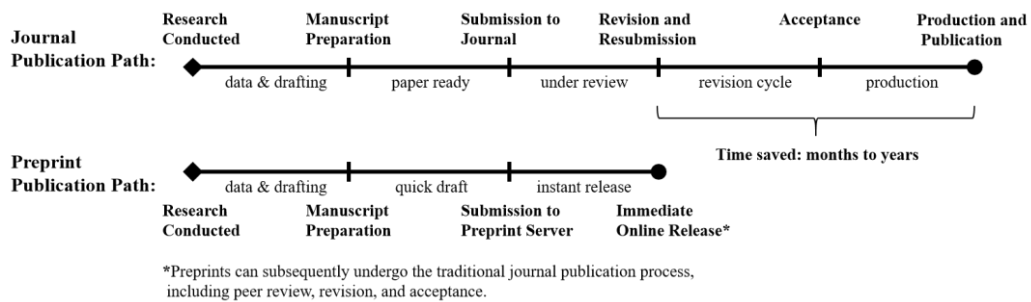

**Fig. A1.** Comparison of journal and preprint publication paths
Note: This figure illustrates the differences between the traditional journal publication path and the preprint publication path. The journal route involves manuscript submission, peer review, multiple revision cycles, acceptance, and production, often requiring months to years before public release. In contrast, the preprint route allows immediate online dissemination after submission to a preprint server, significantly reducing the time lag between research completion and public availability. Preprints can subsequently undergo the traditional journal process, including peer review and acceptance.

**Table A. 1**

Exogeneity test of the visa-free policy.

|  | (1) | (2) |
|---|---|---|
|  | Visa-free | Visa-free |
| SCI/SSCI publications | -0.012 |  |
|  | (0.019) |  |
| Patent applications |  | 0.001 |
|  |  | (0.016) |
| Country FE | Y | Y |
| Month FE | Y | Y |
| Observations | 1904 | 1904 |
| R-squared | 0.606 | 0.606 |

Note: This table tests the exogeneity of the inclusion of countries in China's unilateral visa-free policy. The independent variables are the log-transformed number of SCI/ESCI publications (Column 1) and invention patent applications (Column 2). The results show that the implementation of the visa-free policy is not significantly correlated with these indicators of research and innovation capacity, suggesting that the policy's selection criteria were not systematically driven by national research output or innovation levels. This supports the assumption of exogeneity in this study. All regressions control for country and month fixed effects. ***, **, * indicate significance at the 1%, 5%, and 10% levels, respectively. Standard errors are reported in parentheses.

## Appendix

See Fig. A1 and Table A. 1.

## Acknowledgments

This work was supported by the Fundamental Research Funds for the Central Universities (DUT24RW303).